\documentclass[useAMS,usenatbib]{mn2e}
\usepackage{epsfig}
\usepackage{graphicx}

\newcommand{\ltappeq}{\raisebox{-0.6ex}{$\,\stackrel{\raisebox{-.2ex}{$\textstyle <$}}{\sim}\,$}}
\newcommand{\gtappeq}{\raisebox{-0.6ex}{$\,\stackrel{\raisebox{-.2ex}{$\textstyle >$}}{\sim}\,$}}

\title[The Intrinsic BALQSO Fraction]
{The Intrinsic Fraction of Broad Absorption Line Quasars}

\author[C. Knigge, S. Scaringi, M.R. Goad, and C.E. Cottis]
{Christian Knigge$^{1}$
\thanks{E-mail: christian@astro.soton.ac.uk},
Simone Scaringi$^{1}$,
Michael R. Goad$^{2}$ and Christopher E. Cottis$^{2}$ \\ $^{1}$Department of
Physics and Astronomy, University of Southampton, Highfield, SO17 1BJ, UK\\
$^{2}$Department of Physics and Astronomy, University of Leicester, University
Road, LE1 7RH, UK}

\begin{document}

\date{}

\pagerange{\pageref{firstpage}--\pageref{lastpage}} \pubyear{2006}

\maketitle

\label{firstpage}

\begin{abstract}

We carefully reconsider the problem of classifying broad absorption
line quasars (BALQSOs) and derive a new, unbiased estimate of the
intrinsic BALQSO fraction from the SDSS DR3 QSO catalogue. We first
show that the distribution of objects selected by the so-called
``absorption index'' (AI) is clearly bimodal in $\log{{\rm AI}}$, with
only one mode corresponding to definite BALQSOs. The surprisingly high
BALQSO fractions that have recently been inferred from AI-based
samples are therefore likely to be overestimated. We then
present two new approaches to the classification problem that are
designed to be more robust than the AI, but also more complete than the
traditional ``balnicity index'' (BI). Both approaches yield {\em
observed} BALQSO fractions around 13.5\%, while a conservative third
approach suggests an upper limit of 18.3\%. Finally, we discuss the
selection biases that affect our observed BALQSO fraction. After
correcting for these biases, we arrive at our final estimate of the
{\em intrinsic} BALQSO fraction.  This is $f_{BALQSO} = 0.17 \pm
0.01~{\rm (stat)}\,\pm0.03~{\rm (sys)}$, with an upper limit of
$f_{BALQSO} \simeq 0.23$. We conclude by pointing out that the bimodality
of the $\log{{\rm AI}}$ distribution may be evidence that the BAL-forming
region has clearly delineated physical boundaries.

\end{abstract}

\begin{keywords}
quasars: absorption lines; methods, statistical, catalogues, surveys
\end{keywords}

\section{Introduction}

Broad absorption line quasars (BALQSOs) are a sub-class of
active galactic nuclei (AGN) that exhibit strong, broad and blue-shifted
spectroscopic absorption features
\citep{foltz90,weymann91,reichard03b}. Most BALQSOs -- the so-called
HiBALs -- only display absorption troughs in certain high-ionisation lines
(e.g. NV~$\lambda$1240\AA,  CIV~$\lambda$1549\AA,
SiIV~$\lambda$1397\AA), but some -- the so-called LoBALS -- also show
absorption in some low-ionisation lines (most notably
MgII~$\lambda$2800\AA). BALQSOs are predominantly radio-quiet
\citep{stocke92,becker,shankar}, 
and there are also subtle differences between their
continuum and emission line properties and those of "normal" (non-BAL) QSOs
\citep{reichard03b}. However, despite these differences, BALQSOs and
non-BAL QSOs appear to be drawn from the same parent population
\citep{reichard03b}. 

The simplest and most promising interpretation of the QSO/BALQSO
dichotomy is in terms of an orientation effect. This fits in well with
unified models, in which orientation is the major factor determining the 
observational appearance of AGN (e.g. Elvis 2000). It also makes sense
physically, since the absorption troughs in BALQSOs have long been
recognised as signatures of fast, large-scale outflows from the
central engines. More specifically, blue-shifted 
absorption is produced when the central continuum and/or emission line source
is viewed through outflowing material that scatters photons out of the
observer's line of sight. If the outflow subtends a solid angle $0 < \Omega <
2\pi$, then both BALQSOs and non-BAL QSOs can be accounted for in this
picture. 
\nocite{elvis}

The powerful outflows we observe in BALQSO are an important example of
AGN feedback. Such feedback is the key ingredient in theoretical attempts
to understand galaxy ``downsizing'' and may also be responsible for
regulating the growth of supermassive black holes, quenching star
formation and setting up the $M_{BH}-\sigma$\ and
$M_{BH}-M_{bulge}$\ relations (e.g. Silk \& Rees 1998; King 2003; di
Matteo, Springel \& Hernquist 2005; Scannapieco, Silk \& Bouwens
2005). However, despite their fundamental importance, the geometry,
kinematics and energetics of BALQSO outflows have remained highly
uncertain.
\nocite{silk,DiMatteo,king,scannapieco}

Perhaps the single most important quantity that can be determined
empirically regarding BALQSOs is their incidence within the overall
QSO population. More specifically, the BALQSO fraction ($f_{
BALQSO}$) is defined as the fraction of QSOs that display BALQSO
absorption features. Its significance derives mainly from the fact
that it allows a simple, geometric interpretation: in the context of
unified schemes, $f_{BALQSO}$ is the covering fraction of BALQSO
outflows.

Until recently, searches for BALQSOs in quasar surveys consistently
reported observed BALQSO fractions around 10\%--15\%
\citep{weymann91,Tolea,hewett03,reichard03b}. It therefore came as
something of a surprise when \cite{trump06} reported a
significantly higher BALQSO fraction of 26\% from
the spectroscopic QSO catalogue associated with the 3rd Data Release
(DR3) of the Sloan Digital Sky Survey (SDSS; Schneider et
al. 2005). 
\nocite{schneider}

There can be little question that the QSO sample on which the Trump et
al. study is based is superior to earlier QSO surveys. However, this
is not the reason for their unusually high estimate of
$f_{BALQSO}$. Instead, Trump et al. argue that the 
"classic" definition of BALQSOs, based on the so-called "balnicity index"
(hereafter, BI) is not appropriate for BALQSO classification
purposes. Instead, they prefer a different statistic, the so-called "absorption
index" (hereafter, AI). The AI is designed to be less strict than the
BI, with the result that a significantly higher fraction of
QSOs are classified as broad absorption line (BAL) objects. In
essence, \cite{trump06} argue that BALs can be both weaker and much narrower
than has previously been supposed. QSOs containing such features would
naturally be excluded from any census involving the classic BI definition.

If Trump et~al. are correct, the covering fraction of BALQSO outflows
must be much larger than has previously been assumed. Indeed, it has
been suggested that their {\em observed} BALQSO fraction of 26\%
implies an {\em intrinsic} BALQSO fraction of  
43\%$\pm$2\% once selection effects are taken into account (Dai,
Shankar \& Sivakoff 2008). This is about twice the 
best previous estimates (22\%$\pm$4\% [Hewett \& Foltz 2003]; 15.9\%
$\pm$1.4\% [Reichard et al. 2003]).
\nocite{dai}

The main goals of the present paper are to take a fresh look at the
metrics used for classifying BALQSOs and to derive a new, robust
estimate of $f_{BALQSO}$. It is worth emphasizing from the outset
that what we wish to accomplish is to identify a distinct
sub-population of QSO of which classic BALQSOs 
(with ${\rm BI} > 0~{\rm km~s^{-1}}$) are just the most obvious representatives. This is an 
important point, because -- as effectively argued by Trump et
al. (2006) -- these classic BALQSOs may just be the tip of the
iceberg. Thus the very term ``broad absorption line quasar'' could be
a mis-nomer, since it is possible that the majority of objects
belonging to this population could in principle exhibit only
weak/narrow absorption features (or even no absorption at
all). It could even turn out that a distinct BALQSOs
sub-population does not exist: QSOs could simply exhibit a 
perfectly continuous and smooth distribution of absorption
characteristics, with classic BALQSOs occupying the arbitrarily
defined extreme tail of this distribution. As we shall see, there is, 
in fact, evidence that BALQSOs do form a distinct sub-population. With
this in mind, we will use the term BALQSO throughout this paper to 
denote members of this sub-population, regardless of whether they are
identified as such by any given metric. The goal, in fact, is to find
ways of quantifying the size of this population in a way that is
simultaneously robust (i.e. does not produce many false positives)
and complete (i.e. does not miss many true members). 

In Section~2, we introduce and compare the widely used AI and BI
metrics for identifying BALQSOs. In Section~3, we show that there is 
clear evidence  
for bimodality in the $\log{{\rm AI}}$\ distributions of Trump et al.'s BALQSO
candidates, with "classic" BALQSOs (with positive BI) preferentially
occupying one mode of the distribution. In Section~4, we present
several concrete examples  of problematic classifications obtained
with {\em both} standard metrics. In Section~5, we present two new
approaches to the classification problem, which are designed to be
more robust than the AI, but more complete than the BI. In Section~6,
we correct the {\em observed} BALQSO fractions produced by our new
approaches for selection effects and obtain our final estimate of
the {\em intrinsic} BALQSO fraction. Finally, in Section~7, we discuss 
our results and present our conclusions.

\section{How broad is broad? Metrics for identifying broad
absorption line quasars} 

The BI \citep{weymann91} was the first quantitative metric used to
identify BALQSOs within QSO surveys. Until the introduction of the AI
(see below), the BI remained the standard way to classify objects as 
BALQSOs. Given a continuum-normalised spectrum in the vicinity of a spectral
line, the BI is defined numerically as
\begin{equation}
{\rm BI}=-\int_{25000}^{3000}\left[1-\frac{f(v)}{0.9}\right]Cdv.\label{eq:1}
\end{equation}
Here, the limits of the integral are in units of km~s$^{-1}$, and $f(v)$\ is the
normalised flux as a function of velocity displacement from line
centre.\footnote{It is worth noting that in the original definition of the BI
by Weymann  et al. (1991), $f(v)$ is normalised relative to the
underlying continuum, whereas other authors, including Trump et
al. (2006), normalize relative to a best-fitting continuum plus
emission line template.} The constant $C=0$\ everywhere, unless the
normalised flux has satisfied $f_c(v) < 0.9$\ continuously for at
least 2000~km~s$^{-1}$; at this point it is switched to $C=1$\ until
$f(v) > 0.9$\ again. Based on this definition, objects are classified
as BALQSOs if their ${\rm BI} >0~{\rm km~s^{-1}}$. 

Physically, the idea behind the BI is to count as BALs only absorption
troughs that are definitely real (hence 
the requirement that $f(v) < 0.9$), definitely broad (hence the demand that
troughs must be broader than 2000~km~s$^{-1}$\ in order to count) and
significantly blue-shifted (hence the lower limit of
3000~km~s$^{-1}$\ on the integral). The 
main attraction of the BI as a classification tool is that it tends to produce
very "clean" BALQSO samples. Indeed, it is hard to imagine a non-BAL QSO being
assigned a positive BI unless its spectrum is either very noisy, suffers from
a misplaced continuum, or has been assigned an erroneous redshift. However,
the conservative nature of the BI also means that BALQSO samples based on it
may be seriously incomplete. There is certainly no compelling reason to think
that somewhat weaker, narrower and/or less-blue-shifted BALs than
recognised by the BI should not exist.

This issue was already recognised by \cite{weymann91} and provided the
motivation for the introduction of the AI, initially by
\nocite{hall02} 
Hall et al. (2002, here purely as a means of identifying systems showing evidence of
absorption). The definition of the AI ultimately adopted by Trump
et al. (2006) is 
\begin{equation}
{\rm AI}=\int_{0}^{29000}\left[1-f(v)\right]C^{\prime} dv,\label{eq:2}
\end{equation}
where $f(v)$ is the normalized flux obtained after dividing the data by
the best-fitting emission-line-plus-continuum QSO template. The
constant $C^{\prime}=1$ in all regions where $f(v) < 0.9$ continuously
for at least 1000~km~s$^{-1}$ and $C^{\prime}=0$ otherwise. Also, only regions
containing at least one data point significantly below the underlying
continuum are included in the calculation. This ensures that only true
absorption features are assigned positive AI. The two key
differences that allow some objects with ${\rm BI}=0~{\rm km~s^{-1}}$ to achieve ${\rm AI} >0~{\rm km~s^{-1}}$
are that (i) the AI includes regions within 3000~km~s$^{-1}$ of line
centre (and also regions beyond 25,000~km~s$^{-1}$), and (ii) the AI
includes objects with much narrower absorption troughs than the BI. 
The remaining difference is associated with
the absence of the factor 0.9 in Equation~\ref{eq:2} (compared to
Equation~\ref{eq:1}). This change was made to the definition of the
AI in order to  allow a clear interpretation: the AI is the combined
equivalent width of all absorption troughs in a given line that are
located blue-wards of line centre, deeper than $0.9$ of the continuum,
and at least 1000~km~s$^{-1}$ wide. 

Note that both the AI and the BI can be sensitive to the type of
spectrum from which they are measured. For example, an 
apparently broad absorption trough in a low-resolution spectrum may
break up into multiple narrow troughs when observed at higher
resolution. Conversely, noise spikes may artificially break up a
single trough, so that a true BAL could be assigned zero AI/BI in a
noisy spectrum. Throughout this paper, we will use Trump et al.'s
(2006) AI/BI estimates for objects in the SDSS DR3 QSO catalogue. The
health warning ``as derived from its SDSS spectrum'' should thus
implicitly be added to the AI/BI estimates we use for each QSO.

It is obvious that if BALQSOs are classified on the basis of the less
restrictive AI, the resulting BALQSO fraction will be higher than if
the BI were used. However, it is not obvious {\em a priori} that 
objects selected solely on the 
basis of having ${\rm AI}>0~{\rm km~s^{-1}}$ (i.e. including those
with ${\rm BI} = 0~{\rm km~s^{-1}}$) constitute a single
population. The problem is that a wide variety of non-BAL absorption features
are commonly seen in QSOs and other AGN. These typically narrower features can
be due to absorption at an intermediate redshift along the line of sight to
the QSO, absorption within the host galaxy, or intrinsic absorption
close to the QSO (including the so-called mini-BALS and associated
absorption features) whose origin remains poorly understood and could
conceivably be linked to the broad absorption lines. It is therefore
extremely difficult to say if any particular QSO containing an 
"intermediate" width absorption trough ($1000~{\rm km~s^{-1}} \ltappeq
\Delta v \ltappeq 3000~{\rm km~s^{-1}}$) should be classified as a 
BALQSO or not. Roughly speaking, the BI metric does not consider {\em
any} such objects to be genuine BALQSOs, whereas the AI metric labels
{\em all} such objects as BALQSOs. In the following section, we will
present statistical evidence that the AI metric, in particular, is far
too permissive in this respect. 

\section{The bimodal log(AI) distribution of AI-selected QSOs}
\label{sec:bimodal}

\begin{figure}
\centering
\includegraphics[width=0.5\textwidth]{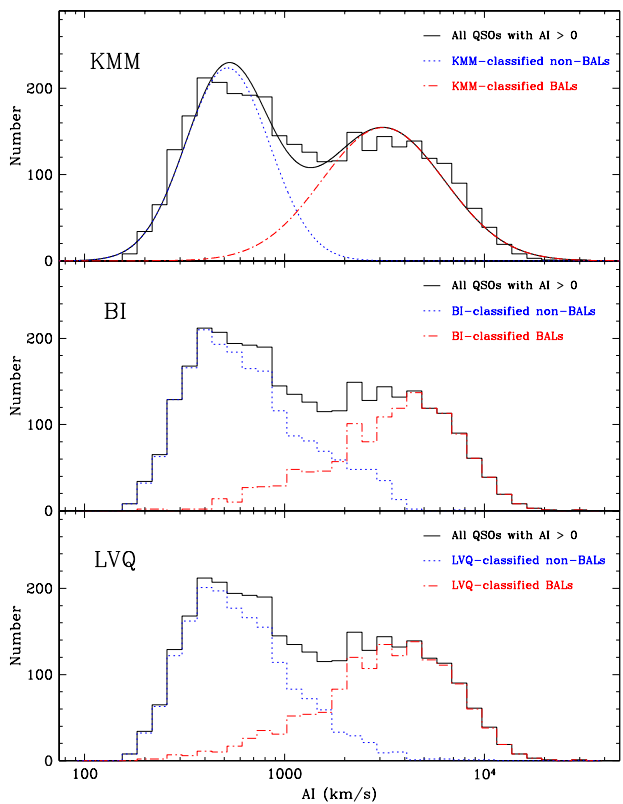}
\caption{The $\log{{\rm AI}}$\ distribution of objects with ${\rm AI} > 0~{\rm km~s^{-1}}$ (black
histograms in all panels). Note the obvious bimodality of this
distribution. {\em Top panel:} The decomposition of the distribution 
suggested by the KMM algorithm. {\em Middle panel:} The decomposition
resulting if the classic balnicity index (BI) is used to classify BALQSOs. {\em
Bottom panel:} The decomposition resulting if a hybrid method
involving learning vector quantization (LVQ) is used to classify
BALQSOs (see Section~5).} 
\label{fig:ais}
\end{figure}

Using the definitions above, Trump et al. (2006) calculated AIs and BIs for
all 11,611 QSOs in the SDSS DR3 sample. In Figure~\ref{fig:ais}, we show
as a black histogram the $\log{{\rm AI}}$\ distribution of the 3182
QSOs with  ${\rm AI} > 0~{\rm km~s^{-1}}$ and in the redshift interval $1.90<z<4.36$ (so as
to contain CIV). This distribution is clearly bimodal, with one peak
near $500$~km~s$^{-1}$ and another around $3000$~km~s$^{-1}$. 

In order to confirm and quantify the bimodality, we have applied the
KMM algorithm of \cite{ashman94}. This effectively compares the
quality of a single Gaussian fit to a distribution to that of a double
Gaussian one. The probability that the
overall $\log{{\rm AI}}$\ distribution is unimodal turns out to be negligible: the
KMM likelihood test ratio statistic (essentially a $\chi^2$) is 590 
for 4 degrees of freedom. This is vastly in excess of the value of 
about 4 one would expect for a unimodal distribution.

The decomposition suggested by KMM is shown in the top panel of
Figure~\ref{fig:ais}. While there is no {\em a priori} reason to
expect the $\log{{\rm AI}}$ distribution to be intrinsically Gaussian
(or double Gaussian), the two normal components provide 
quite a reasonable description of the distribution. More
specifically,  KMM suggests that the low-AI group contributes 49.9\%\
of the total ${\rm AI} > 0~{\rm km~s^{-1}}$ population and is centered on ${\rm AI} \simeq
500~\rm{km~s^{-1}}$ with $\sigma \simeq 0.2$~dex; the high-AI 
group contributes 50.1\% and is centered on ${\rm AI} \simeq 3000~
\rm{km~s^{-1}}$ with $\sigma \simeq 0.3$~dex. 

In the middle panel of Figure~\ref{fig:ais}, we also show the AI
distributions of all objects with ${\rm BI} >0~{\rm km~s^{-1}}$ (red histogram) and of all
quasars with ${\rm BI}=0~{\rm km~s^{-1}}$ but ${\rm AI}>0~{\rm
  km~s^{-1}}$ (blue histogram). This shows that the 
two modes exhibited by the AI-selected quasar population correspond
fairly closely to ``classic'' BALQSOs (high-AI mode; ${\rm BI}> 0~{\rm km~s^{-1}}$), on the
one hand, and newly added objects (low-AI mode; ${\rm BI} = 0~{\rm km~s^{-1}}$), on the
other. The BI metric classifies 41.2\% of the ${\rm AI} >0~{\rm km~s^{-1}}$ objects as 
BALQSOs. In general, the match of the KMM-suggested groups to the
${\rm BI}=0~{\rm km~s^{-1}}$ and ${\rm BI} > 0~{\rm km~s^{-1}}$ groups is good, except near the overlap
region. The KMM decomposition suggests that the BI-selected sample may 
be seriously incomplete in this regime.

It should be acknowledged at this point that ``bimodality'' turns out 
to be a surprisingly slippery concept on closer examination. In
particular, the number of modes in a distribution over a certain 
variable is not always invariant under simple transformations of
that variable. This explains why the bimodality in the
$\log{{\rm AI}}$\ distribution was not noticed by Trump et al. (2006), who
only inspected the (linear) ${\rm AI}$\ distribution. As it turns out, that
distribution is, in fact, unimodal.

\begin{figure*}
\centering
\includegraphics[width=0.65\textwidth,angle=270]{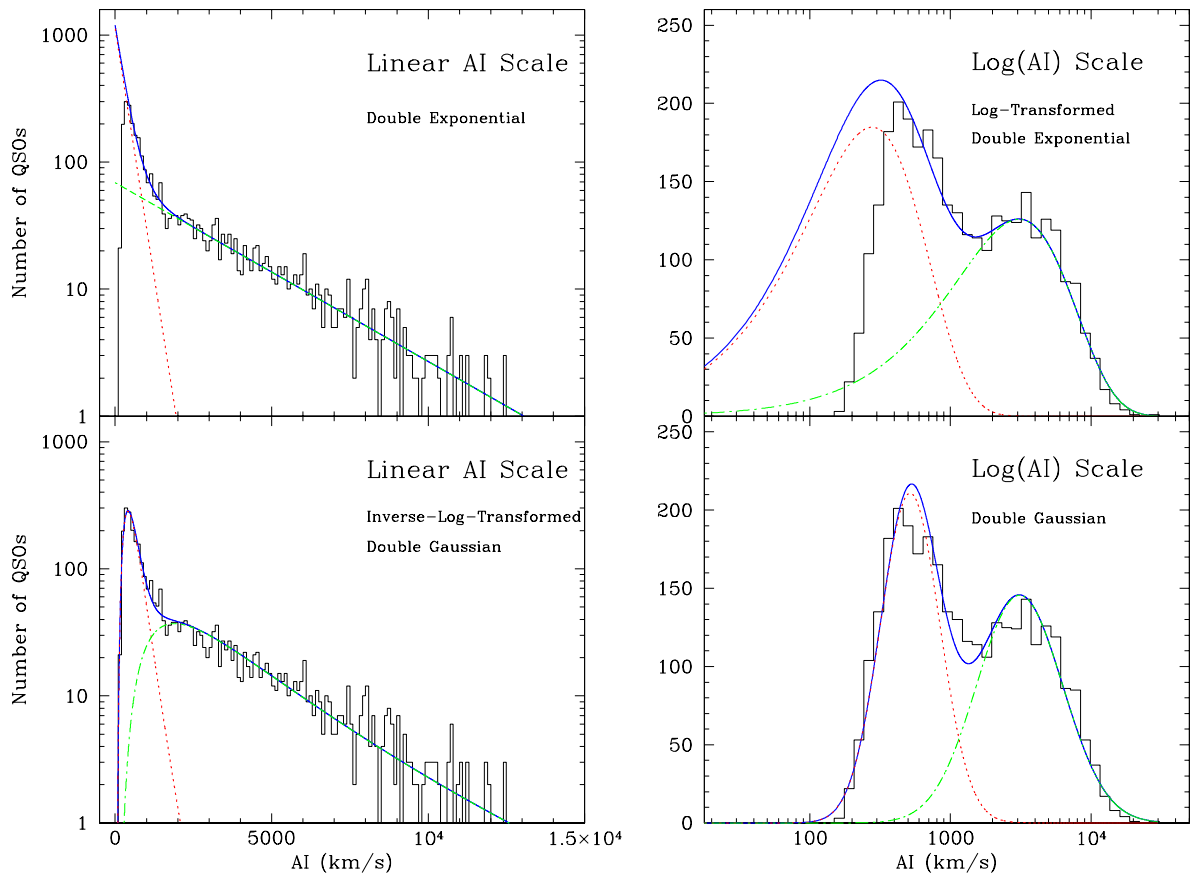}
\caption{Comparison of the ${\rm AI}$\ distribution (left panels) and the
$\log{{\rm AI}}$\ distribution (right panels) of objects with ${\rm AI} > 0~{\rm km~s^{-1}}$. Note that 
only the $\log{{\rm AI}}$\ distribution is bimodal, but that the
${\rm AI}$\ distribution exhibits two distinct characteristic scale lengths
in the low-AI and high-AI regimes. Thus both types of distribution
provide evidence of two distinct sub-populations, each of which
dominates in one of these regimes. In the top panels, we also show a
maximum likelihood fit to the AI-distribution above ${\rm AI} = 500~{\rm
km~s^{-1}}$\ with a double exponential model. In the bottom panels, we
again show the KMM-decomposition from Figure~\ref{fig:ais}, which
corresponds to a double Gaussian in $\log{{\rm AI}}$ (and a double log-normal distribution in AI).} 
\label{fig:ais2}
\end{figure*}

So does the bimodal $\log{{\rm AI}}$\ distribution actually provide evidence
for two distinct QSO sub-populations? It does, because not every
unimodal distribution can be transformed into a bimodal one via a
logarithmic transformation. Thus while the concept of bimodality should
perhaps be replaced by that of ``bimodalizibility''(Wyszomirski
1992)\nocite{wys}, it remains true that 
distinct sub-populations are the most obvious way of producing such
bimodalizable distributions. Indeed, in our case, the evidence for two
distinct QSO sub-populations can be seen even in the linear
${\rm AI}$\ distribution. In Figure~\ref{fig:ais2}, we compare the ${\rm AI}$\ and
$\log{{\rm AI}}$\ distribution directly. Even though the ${\rm AI}$\ distribution
is unimodal, it is obvious that the characteristic scale on which the
distribution drops off changes abruptly at around ${\rm AI} \simeq 1700~{\rm
  km~s^{-1}}$, which coincides with the dip between the two modes of
the $\log{{\rm AI}}$\ distribution. We thus believe that the evidence for
two distinct sub-populations in the overall distribution is
robust.\footnote{Since our paper was accepted, Nestor, Hamann \&
Rodriguez Hidalgo (2008) have also found an excess of strong
absorbers in a study focused mainly on relatively narrow C~{sc iv}
absorption line systems (see their Figure~9). We suspect this excess
may be directly associated with the high-AI, BALQSO mode of the
$\log{AI}$\ distribution.}

Inspection of the linear ${\rm AI}$\ distribution suggests that, beyond its
mode at around ${\rm AI} \simeq 400~{\rm km~s^{-1}}$ (which corresponds to
the low-AI mode of the $\log{{\rm AI}}$\ distribution), the drop-off in 
each of the two distinct regimes is roughly exponential. As shown in
Figure~\ref{fig:ais2}, we have therefore fit a double exponential
model to this distribution for ${\rm AI} > 500~{\rm km~s^{-1}}$. Note that,
as expected, this unimodal two-population model for the
${\rm AI}$-distribution produces a bimodal distribution in
$\log{{\rm AI}}$\ (Figure~\ref{fig:ais2}, top right panel). Since this
double-exponential model imposes no low-AI cut-off at all on the
sub-population that dominates at high-AIs, it allows us to set a
useful upper limit on the size of this population (see 
Sections~\ref{sec:exp} and \ref{sec:alltogethernow}).

%\footnote{Since we treat $\log{AI}$ as the independent
%variable, the distributions we test with KMM are actually log-normal in
%AI. It does not make sense to apply KMM with AI itself
%as the independent variable, since the linear AI distribution covers
%an enormous dynamic range and is
%tremendously skewed (see Trump et al. 2006).} 
%This 
%distribution was actually shown by \cite{trump06}, but on a linear AI
%scale and only up to $AI = 4000$. We believe that this is the
%main reason why the bimodality introduced by the AI definition has so
%far passed unnoticed.

\section{Beyond statistics: representative spectra across the AI/BI 
parameter space}  
\label{sec:beyond}

It is important to relate the statistical results of the previous
section to specific spectral properties of individual QSOs. What type
of objects do we select when we apply AI and/or BI metrics, and what
type of spectra correspond to different combinations of AI and BI?

\begin{figure*}
\centering
\includegraphics[width=\textwidth]{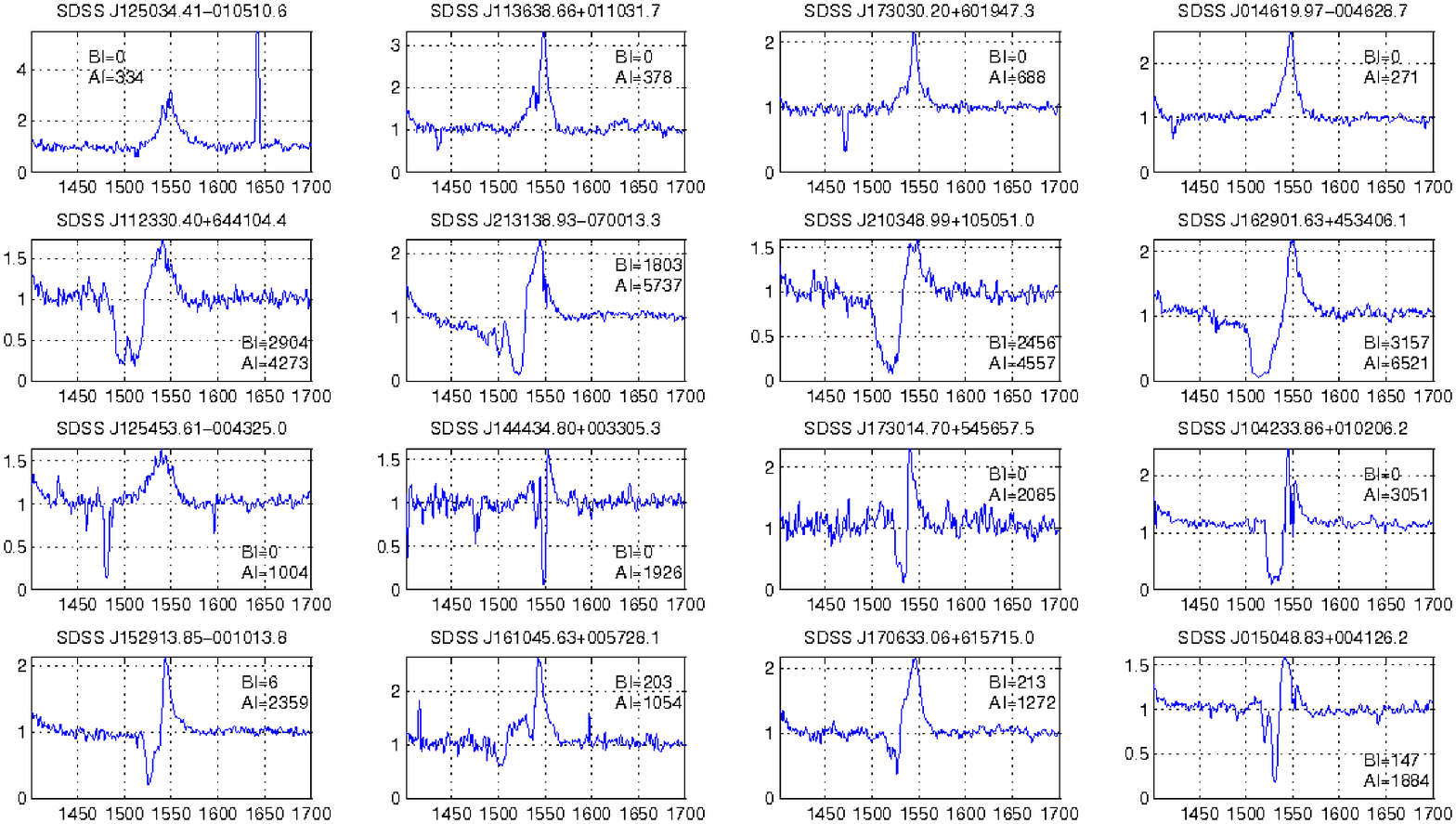}
\caption{Representative spectra for various parts of the AI/BI
parameter space. Each row corresponds to objects from a distinct
region of this parameter space. {\em Top row:} Objects belonging to
the low-AI mode in Figure~1 (${\rm BI}=0~{\rm km~s^{-1}}$; ${\rm AI} \simeq
500$). {\em Second Row (From Top):} Objects belonging to the high-AI
mode in Figure~1 (${\rm BI}>0~{\rm km~s^{-1}}$;  
${\rm AI} \simeq 5000$). {\em Third Row (From Top):} ${\rm BI}=0~{\rm
  km~s^{-1}}$ objects belonging to the overlap 
region in Figure~1 (${\rm AI} \simeq 2000~{\rm
  km~s^{-1}}$). {\em Bottom Row:} 
${\rm BI} > 0~{\rm km~s^{-1}}$ objects belonging to the overlap region in
Figure~1 (${\rm AI} \simeq 2000~{\rm km~s^{-1}}$).} 
\label{fig:specs}
\end{figure*}

The top row in Figure~\ref{fig:specs} shows the C~{\sc iv} line
profiles of four QSOs belonging to low-AI mode in Figure~\ref{fig:ais} (${\rm AI} \sim
500$; ${\rm BI}=0~{\rm km~s^{-1}}$). To our eyes, none of these QSOs 
appear to be genuine BALs.\footnote{It should be acknowledged, however,
that our organic neural networks have also been trained primarily on
{\em BI-selected} BALQSOs.} We have visually inspected 
the majority of similar objects and find that the same is true for
most of them. Objects with ${\rm AI}>0~{\rm km~s^{-1}}$ and ${\rm
  BI}=0~{\rm km~s^{-1}}$ comprise about half 
of the population with ${\rm AI} > 0~{\rm km~s^{-1}}$, so this
population is certainly not representative of ``classic'' BALQSOs. 

The second row from the top in Figure~\ref{fig:specs} shows objects
selected from the high-AI mode in Figure~\ref{fig:ais} (${\rm AI} \simeq 3000$; ${\rm BI} >
0~{\rm km~s^{-1}}$). As expected, all exhibit the strong and broad absorption features
that are characteristic of ``classic'' BALQSOs.

The third row from the top in Figure~\ref{fig:specs} shows a selection
of ${\rm BI}=0~{\rm km~s^{-1}}$ objects from the overlap region in Figure~\ref{fig:ais} (${\rm AI}
\simeq 1000 - 3000~{\rm km~s^{-1}}$). It is immediately clear that these 
intermediate-width absorption line objects can indeed be difficult
to classify with confidence. However, we have also included in this
row two objects (SDSS J1730 and SDSS J1042) that appear to be genuine
BALQSOs that have been missed by the BI. 

The bottom row in Figure~\ref{fig:specs} shows more objects 
from the overlap region in Figure~\ref{fig:ais} (${\rm AI} \simeq 1000 -
3000~{\rm km~s^{-1}}$), but now with ${\rm BI} > 0~{\rm km~s^{-1}}$. While these are clearly harder to
classify than those in the high-AI mode, we think the BI has done a
good job of assigning these objects to the BALQSO class.

In our view, the results of the previous and present sections imply
that, although not perfect, the BI is a better metric for BALQSO
identification than the AI. The majority of objects with positive BI
are clearly genuine BALQSOs, but the same cannot be 
said with any confidence of objects classified solely on the basis of
positive AI. The AI is certainly very good at finding absorbing
systems, including essentially {\em all} BALQSOs. However, the
spectroscopic properties of objects with ${\rm BI}=0~{\rm km~s^{-1}}$ but AI $> 0~{\rm km~s^{-1}}$ , as well
as the bimodality in the AI $> 0~{\rm km~s^{-1}}$ population, suggest that purely
AI-selected BALQSO samples will be strongly contaminated by objects
with properties that are clearly distinct from those of ``classic''
BALQSOs (c.f. Ganguly et al. 2007). 
\nocite{ganguly1,ganguly2}

The fact remains, however, that BI-selected BALQSO samples may
themselves be seriously incomplete. In Section~3, we showed that the BI
criterion selects 41.2\% of QSOs with ${\rm AI} > 0~{\rm km~s^{-1}}$ as BALQSOs, whereas the 
KMM decomposition of the $\log{{\rm AI}}$\ distribution implies a significantly higher
percentage of 50.1\%. Similarly, we have now found specific examples
of QSOs with ${\rm BI}=0~{\rm km~s^{-1}}$ that, visually, would seem to be excellent BALQSO
candidates (e.g. SDSS J1730 and SDSS J1042 in
Figure~\ref{fig:specs}). None of this should come as a surprise. As
discussed in Section~2, there
is simply no physical reason to expect that all genuine BALQSOs should
have C~{\sc iv} absorption troughs that extend for at least
2000~${\rm km~s}^{-1}$ beyond the arbitrary 3000~${\rm km~s}^{-1}$ starting point
adopted in the definition of the BI.

We conclude that BALQSO fractions derived from AI-selected samples are
strong overestimates, whereas those derived from BI-selected samples
are at least mild underestimates. In the following section, we will
use two new methods to determine observed BALQSO fractions that are
more robust than AI-based estimates and more complete than BI-based
ones.

\section{The observed BALQSO fraction in SDSS DR3}

The fundamental problem with simple metrics such as the AI and the BI is
their rigidity. For example, the BI will firmly reject an object with
an absorption trough whose width is marginally less than
2000~${\rm km~s}^{-1}$, even if this trough looks virtually
indistinguishable from many objects that the BI {\em does} classify
as BALQSOs. One way to avoid this incompleteness is to relax the
classification criteria, but this incurs the danger of producing many
false positives. This is what appears to have happened in the switch
from the BI to the AI.

In order to overcome these problems, we have used two new approaches
to estimate the observed BALQSO fraction in SDSS DR3. The first
approach is based directly on the KMM-decomposition of the AI
distribution in Figure~\ref{fig:ais}, whereas the second approach is a
hybrid method that employs a BI-trained neural network algorithm --
learning-vector quantization (LVQ) -- to flag potentially
mis-classified objects for visual inspection. We also use a third
approach -- a decomposition based on the double-exponential model for
the ${\rm AI}$\ distribution described in Section~\ref{sec:bimodal} --
to estimate an upper limit on the observed BALQSO fraction. The feature
common to all three approaches is that they are fundamentally more
flexible than the AI or BI metrics. 

\subsection{KMM-based decomposition}
\label{sec:kmm}

The KMM-based approach is straightforward. As discussed in Section~3
and shown in Figure~\ref{fig:ais} (top panel), the $\log{{\rm AI}}$\ distribution of
QSOs with ${\rm AI}>0~{\rm km~s^{-1}}$ can be decomposed fairly cleanly into two
Gaussian components. This decomposition can be used immediatley to assign a
probability to each object of belonging to one or the other group. The
KMM algorithm we have used provides these probabilities automatically
for each object. A raw, observed BALQSO fraction can therefore be
estimated from this decomposition as 
\begin{equation}
f_{BALQSO} = \frac{1}{N_{QSO}} \sum_{i=1}^{N_{QSO}} P_{i,BALQSO}
\label{eq:kmm}
\end{equation}
where $P_{i,BALQSO}$ is the KMM-assigned probability that quasar $i$
is a BALQSO (i.e. that it belongs to the high-AI mode of the
distribution). 

The main weakness of this method is that it assumes the 
KMM-decomposition to be correct. This is almost certainly not true in 
detail. Just as there is no reason to think that every BALQSO trough
is at least 2000~${\rm km~s}^{-1}$\ wide, there is no {\em a priori} reason
to assume that the $\log{{\rm AI}}$\ distribution of BALQSOs is exactly
Gaussian. However, Figure~\ref{fig:ais} suggests that a Gaussian
distribution may be quite a good approximation to the true
$\log{{\rm AI}}$\ distribution of BALQSOs. The great strength of the decomposition
approach is that it provides a very complete statistical census of
BALQSOs (subject to its underlying assumption).

Applying this method to the full DR3 QSO sample in the redshift range
$1.90 < z < 4.36$ yields an observed BALQSO fraction of 13.7\% 
$\pm$0.3\% (where the error only accounts for Poisson
statistics). Note that this observed global fraction is still subject
to selection biases. These are dealt with in Section~6. 

\subsection{A hybrid method using learning vector quantization} 
\label{sec:lvq}

In our second approach, we use a hybrid method to classify
BALQSOs. Starting with a BI-based classification, we use a machine
learning algorithm called  
Learning Vector Quantization (LVQ) to identify objects that might have
been misclassified by the BI. All such objects are then inspected and
classified visually. We will refer to this hybrid method as
``LVQ-based'' throughout this paper. However, it should be kept in mind
that we do not use LVQ as a stand-alone BALQSO classifier, but as part
of a process involving the BI, LVQ and visual inspection. 

LVQ was originally devised by \cite{kohonen01} and uses a neural
network to assign new input data 
to pre-defined classes. LVQ is a particulary simple supervised neural
network, in which each neuron is simply tagged as belonging to a 
particular class. The basic idea 
behind LVQ is that, through training, each neuron should come to
represent a characteristic type of object within its class. New inputs
can then be assigned to classes on the basis of maximum similarity to
a particular neuron. 

In our case, the relevant classes are BALs vs non-BALs, and the
data are continuum-normalised QSO spectra between
$\lambda\lambda$1400--1700 \AA\ (spanning the C~{\sc iv}
line). Normalization is performed exactly as 
in \cite{north06}, and the measure of similarity we use when comparing 
spectra and neurons is the Euclidean distance between them (i.e. the
mean rms residual). Note that we allow for redshift errors in all
comparisons.

We use 800 QSOs as our training set, with 400 ${\rm BI}>~ 0 {\rm km~s^{-1}}$ objects initially
representing the BALQSOs and 400 ${\rm BI}=0~{\rm km~s^{-1}}$ objects initially representing
the non-BAL QSOs. The network is then iteratively trained to classify
the objects in the training set in line with their input
classifications. Even though these input classifications are purely
BI-based, the converged network already
manages to classify some ${\rm BI} = 0~{\rm km~s^{-1}}$ objects in the training set as
likely BALQSOs (and some ${\rm BI} > 0~{\rm km~s^{-1}}$ objects as likely non-BAL
QSOs). This is possible 
because the network classifications are based on spectral similarity, 
not on the BI itself. In order to re-enforce this feature of the
network, we inspect all of the ``misclassified'' objects in the
training set visually and retag them if appropriate. We then carry out
a full second training run, where the training set now includes ${\rm BI}=0~{\rm km~s^{-1}}$
objects explicitly tagged as BALQSOs (and vice versa). The converged
network produced by this second training run is our final LVQ machine
classifier. All 11,611 DR3 QSOs in the relevant redshift range are
passed to this network, resulting in an LVQ classification for each of
them.

As already noted above, we do not use LVQ as a stand-alone BALQSO
classifier, but as a tool to flag borderline cases where the 
LVQ and BI classifications disagree. All such cases are then inspected 
and classified visually, and the visual classification is adopted as
final. In practice LVQ classified 524 ${\rm BI} =
0~{\rm km~s^{-1}}$ 
objects as BALQSOs, of which 334 were also classified as BALQSOs
visually. Thus LVQ was 
quite good at identifying 
${\rm BI} = 0~{\rm km~s^{-1}}$ BALQSOs. However, LVQ also classified 383 objects with ${\rm BI} >
0~{\rm km~s^{-1}}$ as non-BAL QSOs, and only 95 of these were also
classified as non-BAL QSOs visually. 
This underlines the importance of the visual inspection
step and justifies our reluctance to use LVQ as a stand-alone BALQSO
classifier. As explained above, whenever we refer to ``LVQ-based''
quantities below, we will always mean quantities calculated on the
basis of the full hybrid method, which uses the BI, LVQ and visual
inspection.

Our LVQ-based method classifies 1,557 of the 11,611 QSOs in our DR3
parent sample as BALQSOs. \footnote{A catalogue providing the
KMM-assigned probabilities and  LVQ-based classifications is
available in electronic form from http://www.astro.soton.ac.uk/~simo .}
The LVQ-based decomposition of ${\rm AI} > 0$ objects into BALQSOs and
non-BAL QSOs is shown in the bottom panel of Figure~\ref{fig:ais}. 
The LVQ-based observed BALQSO fraction is 13.4\%~$\pm$~0.3\%, which
is consistent with the KMM-based estimate. 

\subsection{Double exponential decomposition}
\label{sec:exp}

Our third and final approach is based on the double exponential model
for the (linear) AI distribution described in
Section~\ref{sec:bimodal} and shown in Figure~\ref{fig:ais2}. If we
associate the exponential that dominates at high-AIs with BALQSOs, we
can use this model to estimate BALQSO fractions in the same way as for
the KMM-based decomposition. It is worth emphasizing that this model
assumes that there is no low-AI cut-off at all in the true BALQSO
population -- even QSOs with no absorption at all can be ``BALQSOs''
in this case. The observed turn-over in the AI-distribution below ${\rm AI}
\simeq 400~{\rm km~s^{-1}}$ must then be due to incompleteness. This
is not entirely unreasonable, since the definition of the AI imposes a
lower limit of $100~{\rm km~s^{-1}}$\ and only counts absorption
troughs that dip below true continuum. 

While this is quite an extreme model in our opinion, it is impossible
to rule out with the present data. We have therefore also estimated an
``observed'' BALQSO fraction on the basis of this double-exponential
decomposition. This effectively provides an upper limit on the BALQSO
fraction. Based on the model shown in Figure~\ref{fig:ais2}, we find
that the exponential dominating at high-AI values corresponds to an
observed BALQSO fraction of 18.3\%~$\pm$~0.4\% (where the errors are
again purely based on Poisson statistics). Note that this estimate
includes QSOs with estimated ${\rm AI} = 0~{\rm km~s^{-1}}$ that are not
part of Trump et al.'s (2006) AI-based BALQSO catalogue. If we
(somewhat arbitrarily) exclude such objects, the observed BALQSO
fraction is 17.2\%~$\pm$~0.4\%. As explained above, we consider these
estimates to be upper limits on the observed BALQSO fraction. It is
therefore worth noting that even the estimate which includes ${\rm AI} =
0~{\rm   km~s^{-1}}$ objects lies substantially below the 26\% BALQSO
fraction suggested by Trump et al. (2006) based on the number of QSOs
with ${\rm AI} >  0~{\rm km~s^{-1}}$. 

\section{The intrinsic BALQSO fraction}

\begin{figure}
\centering
\includegraphics[angle=270,width=0.5\textwidth]{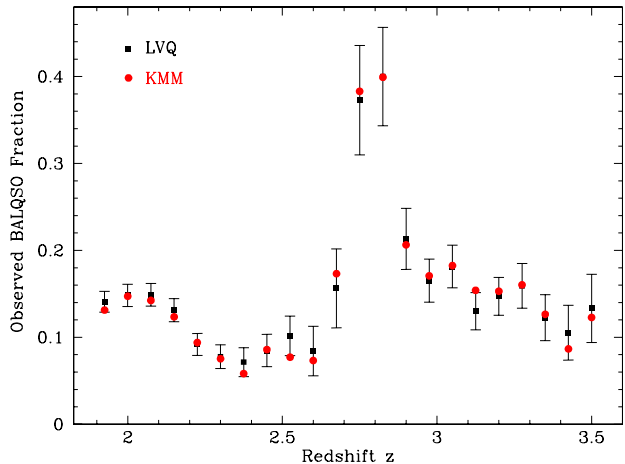}
\caption{The redshift distribution of the {\em observed} BALQSO
fraction. Red points correspond to the fractions determined with the
KMM-based approach (see Section~\protect\ref{sec:kmm}); black points
correspond to the fractions obtained from the LVQ-based approach (see
Section~\protect\ref{sec:lvq}). No correction for selection effects
has been applied to these fractions. Error bars on the KMM fractions
have been suppressed for clarity, but are always similar to the LVQ
ones.}  
\label{fig:rawfrac}
\end{figure}

The observed BALQSO fractions we have derived in the previous section
do not provide a fair measure of the {\em intrinsic} incidence of BALs
within the QSO population. This is because the SDSS QSO sample suffers
from a variety of selection effects that affect BALQSOs differently
from non-BAL QSOs. The impact of the resulting biases can be seen in
Figure~\ref{fig:rawfrac}, which shows that the observed BALQSO
fractions depend strongly on redshift. As we shall see, this redshift
dependence is mainly due to selection effects (c.f. Reichard et
al. 2003).

In the following subsections, we first construct a more homogenously
selected QSO sample and then correct the observed BALQSO 
fraction derived from it for colour-, magnitude- and
redshift-dependent biases. Finally, we put all of these results
together to produce an unbiased estimate of the intrinsic BALQSO
fraction.

\subsection{A homogenous QSO parent sample}

The SDSS DR3 QSO catalogue contains objects selected via a variety of
selection criteria (Schneider et al. 2005). We therefore create a more
homogenous QSO sample by retaining only those objects that were (or
would have been) selected for spectroscopic follow-up by the final QSO
targeting algorithm (as described by Richards et al. 2002). This
leaves us with 7,487 QSOs (out of 11,611) in our redshift range. The 
observed BALQSO fractions in this homogenous sample are 14.0\%~$\pm$~0.4\%
(KMM-based) or 14.1\%~$\pm$~0.4\% (LVQ-based), but still exhibit a
strong redshift dependence due to selection effects.
\nocite{Richards}

The SDSS QSO selection algorithm actually  consists of two parallel
strands, one aimed at creating a  ``main'' QSO sample, the other 
aimed specifically at finding high redshift QSOs.
\footnote{Strictly speaking, there is also a third strand, since
objects with FIRST radio counterparts are also preferentially 
targetted. However, in order to correct for optical colour- and 
magnitude-dependent biases, we need a sample with rigorous optical
selection criteria. We therefore do not include QSOs targetted solely
on the basis of radio emission in our homogenous QSO sample.} 
The two strands use different limiting $i^{\prime}$-magnitudes and
colour selection criteria, which must be taken into account when
dealing with the resulting selection biases. Of the 7,487 objects in our
homogenous sample, 5134 would have been selected by the main sample
selection criteria and 4145 by the high-redshift QSO selection
criteria (1792 QSOs satisfied both sets of criteria). 

\subsection{Limiting-magnitude bias}
\label{sec:kcorr}

There are two reasons why a magnitude cut may affect BALQSOs
differently from non-BAL QSOs. First, BAL troughs may be redshifted 
into the bandpass where the magnitude cut is applied, causing BALQSOs
to appear fainter than otherwise identical non-BAL QSOs. Second, the
{\em continuum} spectral energy distributions (SEDs) of BALQSOs 
are reddened with respect to those of non-BAL QSOs. As already shown
by Reichard et al. (2003), the form of this reddening is consistent
with extinction by SMC-like dust. This again means that BALQSOs will
be fainter than otherwise similar non-BAL QSOs. The consequence of
these effects is that any magnitude cut will disproportionately remove
BALQSOs from the sample.

\begin{figure}
\centering
\includegraphics[width=0.5\textwidth]{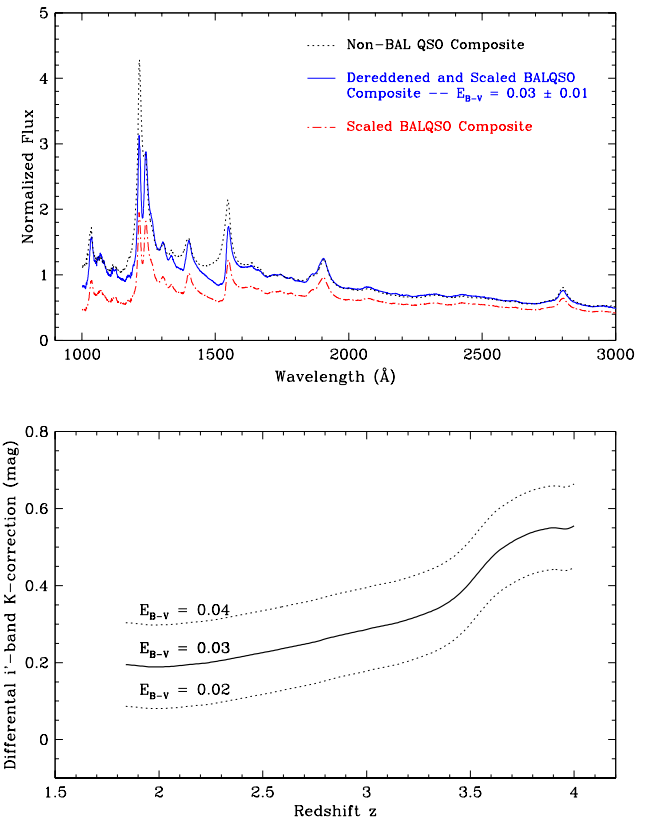}
\caption{LVQ-based BALQSO and non-BAL QSO composites and the
corresponding K-correction in the $i^{\prime}$-band. {\em Top Panel:} The
blue line shows the BALQSO composite after dereddening and scaling it 
to optimally match the non-BAL QSO composite (black line). The red
line shows the original BALQSO spectrum, but scaled so that its
normalization relative to the non-BAL spectrum is in line with the
difference in reddening/extinction between them. {\em Bottom Panel:}
Each line shows the redshift-dependent $i^{\prime}$-band magnitude
difference between the correctly scaled BALQSO and non-BAL QSO
composites, for a particular assumed value of the differential
reddening/extinction between them. The solid dark line corresponds to 
our preferred reddening estimate; the dotted thin lines as based on 
our estimate of the uncertainty on this.} 
\label{fig:kcorr}
\end{figure}

In order to correct for this, we first construct BALQSO and
non-BAL QSO composites and estimate the difference in
reddening/extinction between them. 
\footnote{The composites used in this section were constructed 
using the LVQ-based samples; the equivalent KMM-based
composites are virtually identical.} Note that we use geometric mean
composites, which ensures that the spectral index and reddening of
each composite corresponds to the arithmetic mean of the spectral
indices and reddening values of the spectra used to construct the
composite (Reichard et al. 2003). The absolute flux densities of the
composites are arbitrary, however, since all individual spectra are
scaled to an average value of unity in a reference wavelength interval
near 1700~\AA. 

As shown in Figure~\ref{fig:kcorr} (top panel), dereddening the BALQSO 
composite by $E(B-V) = 0.03 \pm 0.01$ and rescaling produces a good
match to the non-BAL QSO composite longward of $\simeq
1600$~\AA\ (i.e. away from any major 
BAL troughs). This is consistent with the findings of Reichard et
al. (2003). We therefore scale the original BALQSO composite so that
its normalization relative to the non-BAL QSO composite is in line
with our estimate of the difference in extinction between them
(Figure~\ref{fig:kcorr}; red line). Finally, we carry out synthetic
photometry to determine the $i^{\prime}$-band 
magnitude difference between BALQSOs and non-BAL QSOs as a function of
redshift. This ``differential K-correction'' is shown in the bottom
panel of Figure~\ref{fig:kcorr}. The sharp upturn around $z \simeq
3.5$ corresponds to the first major BAL trough (C~{\sc iv}~1550~\AA)
being red-shifted into the $i^{\prime}$-band. At the lower redshifts
we will mostly be interested in below, the K-correction is only due to
extinction.  

We can now estimate a corrected BALQSO fraction in any redshift bin as
\footnotesize
\begin{equation}
f_{BALQSO} = \frac{N_{BALQSO}}{N_{BALQSO} + 
N_{non-BAL\, QSO}(i^{\prime} < \left[i^{\prime}_{lim} - \Delta
  i^{\prime}(z)\right])},
\label{eq:kcorr}
\end{equation}
\normalsize
where $i^{\prime}_{lim}$\ is the limiting magnitude imposed by the
selection algorithm ($i^{\prime}_{lim} = 19.1$\ for any QSO selected 
only via the main sample strand; $i^{\prime}_{lim} = 20.2$\ for QSOs
identified by the high-z colour selection). The quantity $\Delta
i^{\prime}(z) > 0$ is the differential K-correction. For
sufficiently narrow redshift bins, this could be approximated as
constant within each bin, but it is just as easy (and more precise) to
calculate the K-correction independently for each non-BAL QSO
according to its exact redshift. Note that $N_{BALQSO}$ and
$N_{non-BAL \, QSO}$
become sums over probabilities when calculating $f_{BALQSO}$ 
from the probabilistically-defined KMM sample (c.f. Equation~\ref{eq:kmm}). 

Our correction for limiting-magnitude bias is similar to that applied
by Hewett \& Foltz (2003). It should produce reasonable 
results, provided that the intrinsic BALQSO and non-BAL QSO luminosity 
functions do not exhibit sharp breaks near the limiting absolute
magnitude in any given redshift bin. One limitation of our approach 
is that it does not account for variations in K-correction associated
with variations in BAL strength. However, BAL troughs only
affect the K-correction beyond $z \gtappeq 3.5$, and in this regime we
also do not  have a reliable correction for colour-selection bias (see
Section~\ref{sec:colcorr} and Figure~\ref{fig:colcorr}). We therefore
simply restrict our attention to lower redshifts, $z \ltappeq 3.5$.

\subsection{Colour-selection bias}
\label{sec:colcorr}

Both the main and high-z strands of the SDSS targeting algorithm
select QSO candidates on the basis of their optical photometric
colours. In both strands, QSO candidates are identified as
outliers from the locus defined by normal stars in the 5-dimensional
SDSS colour space (u$^{\prime}$g$^{\prime}$r$^{\prime}$i$^{\prime}$z$^{\prime}$).  
The completeness of the resulting QSO samples is a function of
redshift, since even a fixed intrinsic SED produces different observed
colours when placed at different redshifts. 

None of this would matter for the derivation of the intrinsic BALQSO
fraction if the SEDs of BALQSOs and non-BAL QSOs were identical (at 
least in a statistical sense). Unfortunately, they are not. First,
whenever a deep BAL trough is shifted into a particular waveband, all
colours involving that band are changed. Second, as discussed in
Section~\ref{sec:kcorr} and shown in Figure~\ref{fig:kcorr}, the {\em
continuum} SEDs of BALQSOs are reddened compared to those of non-BAL
QSOs. The upshot of these colour differences is that the efficiency of
the SDSS QSO selection algorithm(s) is not the same for BALQSOs as for
non-BAL QSOs. 

\begin{figure}
\centering
\includegraphics[width=0.5\textwidth]{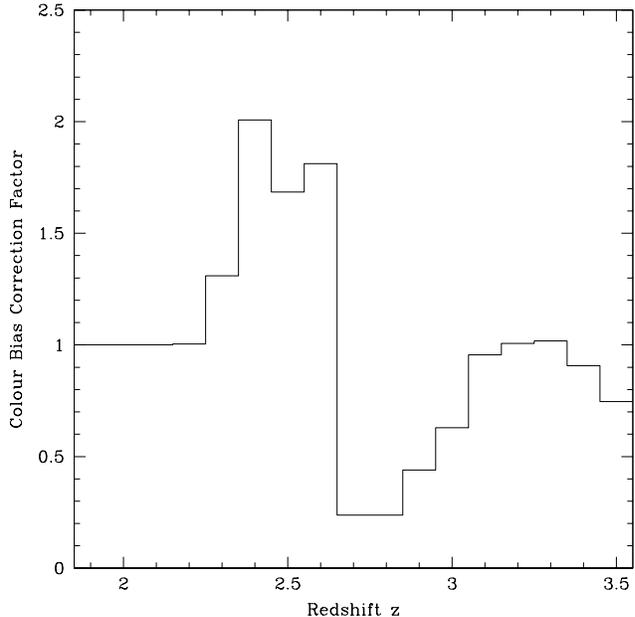}
\caption{The redshift-dependent correction factor for colour-selection
bias. This was derived by Reichard et al. (2003; see their Fig. 10),
and needs to be applied to the observed BALQSO fraction.}
\label{fig:colcorr}
\end{figure}

Fortunately, Reichard et al. (2003) have already derived a
redshift-dependent correction factor that can be applied to the 
observed BALQSO fraction to account for this colour-selection bias. In
order to determine this correction, Reichard et al. created large sets of
simulated QSO and BALQSO colours and passed both through the SDSS QSO
selection algorithm. The resulting correction factor is shown as
a function of redshift in Figure~\ref{fig:colcorr}.

Three key points should be noted regarding this correction for 
colour-selection bias (see Reichard et al. 2003 for a full
discussion). First, it is only approximate. One important
limitation 
is that all of the simulated BALQSO colours used to derive the
correction factor were based on the colour differences between a
HiBALQSO composite and an average QSO composite. Thus variations in
BALQSOs colours arising from the range of observed BAL strengths are 
not properly accounted for. It should also be kept in mind that the
HiBALQSO composite used by Reichard et al. was based on a different
definition of what constitutes a BALQSO than the LVQ- or KMM-based
definition used here. Second, the correction factor is
significantly greater than unity near $z \simeq 2.5$, but much less
than unity near $z \simeq 2.8$. Thus the colour-selection bias causes
BALQSOs to be under-represented around $z \simeq 2.5$, but
over-represented around $z \simeq 2.8$ (this explains the spike at
this redshift in Figure~\ref{fig:rawfrac}). Third, the correction
factor is close to unity for $z \ltappeq 2.2$ and $3.0 \ltappeq z
\ltappeq 3.4$. These  redshift ranges are thus optimal for estimating
the intrinsic BALQSO fraction.

\subsection{Putting it all together}
\label{sec:alltogethernow}

Let us summarize all of the steps we have taken so far. First, we
assigned a BALQSO or non-BAL QSO classification to every object in the 
redshift range $1.90 < z < 4.36$ in the SDSS DR3 QSO
catalog.\footnote{In the case of KMM, we assign a BALQSO probability.} 
Next, we
created a more homogenous sample by removing all objects that were not
selected by  
the SDSS QSO targetting algorithm. We then accounted
for limiting-magnitude bias by removing every non-BAL QSOs that is 
fainter than the effective (dereddened) magnitude limit for a BALQSO
at the same redshift. Finally, we applied a redshift-dependent
correction for the colour-selection bias imposed by the QSO targetting
algorithm.

\begin{figure}
\centering
\includegraphics[angle=270,width=0.5\textwidth]{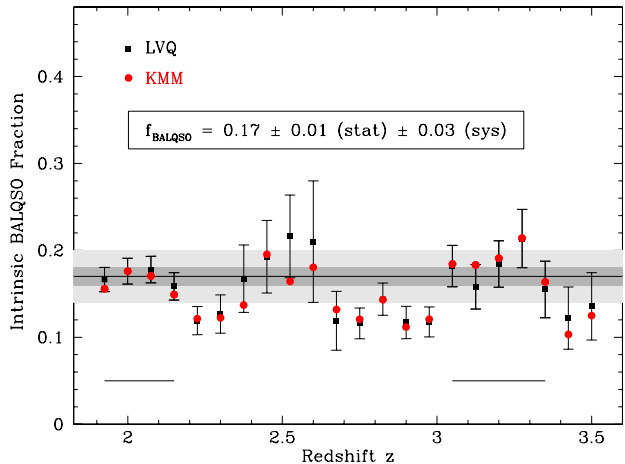}
\caption{The redshift distribution of the {\em intrinsic} BALQSO
fraction (after correcting for selection effects). Red points
correspond to the fractions determined with the KMM-based approach
(see Section~\protect\ref{sec:kmm}); black points correspond to the
fractions obtained from the LVQ-based approach (see
Section~\protect\ref{sec:lvq}). Note that the redshift dependence of
the {\em intrinsic} fractions is markedly reduced compared to that of
the {\em observed} fractions
(c.f. Figure~\protect\ref{fig:rawfrac}) and that the residual
undulations are strongly correlated with the correction factor
for colour-selection bias (Figure~\protect\ref{fig:colcorr}). The
horizontal lines near the bottom of the plot marks redshift ranges
where colour-selection bias is negligible. Our final estimate of the
intrinsic intrinsic BALQSO fraction is derived from only those regions
and is shown the solid horizontal line. The dark (light) shaded
regions correspond to our estimate of the statistical (systematic)
uncertainty on this number.}
\label{fig:balfrac}
\end{figure}

The final product of all these steps -- and the main result of this
paper -- is the intrinsic BALQSO fraction plotted in
Figure~\ref{fig:balfrac}. Two key points are worth noting from 
this straightaway. First, the agreement between the KMM- and LVQ-based
BALQSO fractions is extremely good across the whole redshift
range. This adds to our confidence that we are measuring the intrinsic
abundance of a consistent class of objects. Second, the intrinsic
BALQSO fractions show much less variability with redshift than the observed
fractions (c.f. Figure~\ref{fig:rawfrac}), although some 
residual ``wiggles'' remain. Comparing Figures~\ref{fig:colcorr} and
\ref{fig:balfrac} immediately suggests that these wiggles are due to an
imperfect correction for colour-selection bias. More specifically,
the redshift dependence of the colour-correction factor is
positively correlated with that of the intrinsic BALQSO fraction, so the 
correction derived by Reichard et al. (2003) appears to be somewhat too
strong at most redshifts. We therefore do not believe that there is
evidence for genuine evolution in $f_{BALQSO}$ with redshift.

As suggested in Section~\ref{sec:colcorr}, we derive our final estimate of the
intrinsic BALQSO fraction from the restricted redshift ranges $1.9 < z
< 2.2$ and $3.0 < z < 3.4$.  These are largely free of
colour-selection bias and produce consistent results. Our best
estimate of the intrinsic BALQSO fraction from  these regions is
$f_{BALQSO} = 0.17 \pm 0.01~{\rm (stat)}\,\pm0.03~{\rm (sys)}$. 
The statistical error here is just due to number
statistics. The systematic error accounts for the uncertainty on the
differential K-correction and for alternative choices in constructing
the parent sample and selecting optimal redshift ranges.

We finally also estimate an upper limit on the intrinsic BALQSO
fraction, based on the double exponential decomposition described in
Section~\ref{sec:exp}. The upper limit on the observed BALQSO fraction
suggested by this decomposition was 18.3\%, approximately 1.35 times
larger than our preferred estimates of 13.7\% (KMM) and 13.4\%
(LVQ). Since there is no evidence for a redshift dependence, we
estimate an upper limit on the intrinsic fraction by applying the
same factor to our best estimate of this fraction. The resulting upper 
limit is then $f_{BALQSO} \simeq 0.23$.

\section{Discussion and Conclusions}

Determining the ``true'' BALQSO fraction is a challenging task. A
large part of the problem is the ambiguity one often encounters when
attempting to classify individual absorption features as BALs or
otherwise. The first goal of the present work has been to shed light
on this classification problem. In this context, we have shown that 
when the recently introduced ``absorption index'' (AI) is used 
to classify BALQSOs, the resulting $\log{{\rm AI}}$\ distribution is clearly 
bimodal. Both modes contain comparable numbers of objects, but only
the high-AI mode is clearly associated with genuine BALQSOs. Thus
recent AI-based estimates of the BALQSO fraction -- 26\% (observed;
Trump et al. 2006) or 43\% (intrinsic; Dai, Shankar \& Sivakoff 2008)
-- are likely to be seriously overestimated. 

However, there are also good reasons to believe that the traditional
``balnicity index'' (BI) produces incomplete BALQSO samples. In order
to make progress, we have therefore used two 
complementary new approaches to derive observed BALQSO fractions. One
is based on a statistical decomposition of the $\log{{\rm AI}}$\ distribution, 
the other is a hybrid method in which a BI-trained neural network
flags likely mis-identifications for visual inspection. Both
approaches yield an observed BALQSO fraction around 13.5\% for the
SDSS DR3 QSO catalog (in the range  $1.90<z<4.36$). This number
should be more reliable than AI-based ones and more complete than
purely BI-based ones. We also estimate an upper limit on the observed
fraction of 18.3\%, based on a decomposition of the AI-distribution
that allows even objects without any absorption to be classified as
BALQSOs.

This observed fraction is still subject to serious selection
effects. We have therefore explained in detail how the observed BALQSO 
fraction can be corrected for colour-, magnitude- and
redshift-dependent selection biases. Along the way, we confirmed
that BALQSOs have redder SEDs than non-BALs, consistent with
extinction by SMC-like dust at a level of $E(B-V) = 0.03 \pm 0.01$.

After applying all corrections, there is no compelling evidence for
redshift evolution in the intrinsic BALQSO fraction. Our final
estimate of the global intrinsic BALQSO fraction is then 
$f_{BALQSO} = 0.17 \pm 0.01~{\rm (stat)}\,\pm0.03~{\rm (sys)}$, with
an upper limit of $f_{BALQSO} \simeq 0.23$. As expected, this is
similar to, but slightly higher than, the BI-based estimates from the
SDSS EDR (Reichard et al. 2003). It is also similar to recent BI-based
estimates (Hewett \& Foltz 2003; Dai, Shankar \& Sivakoff 2008) and
consistent with the BALQSO fraction measured by Maddox et al. (2008)
from a K-band selected QSO sample. \nocite{maddox}

In closing, we would like to comment on the relationship
between BALQSOs and what might be called ``absorption line
QSOs''(ALQSOs; this includes all objects displaying some form of
absorption, such as BALs, mini-BALs, associated  absorption features,
narrow absorption 
lines...). Based primarily on the bimodality of the
$\log{{\rm AI}}$\ distribution, we have argued 
throughout this paper that BALQSOs represent a phenemenologically
distinct class amongst the ALQSOs. However, this does {\em not} imply
that BALs and other absorption features must be produced in physically
distinct line-forming regions. After all, orientation effects alone can
dramatically alter the appearance of lines formed in non-spherical
outflows from accretion disks (see, for example, Hamann, Korista \&
Morris [1993], Murray et al. [1995], or, in a different context, Knigge et 
al. [1995], Long \& Knigge [2002]). Indeed, in the QSO unification scheme
of Elvis (2000), both broad and narrow absorption lines are explicitly
assumed to be formed in the same disk wind. In our view, it is likely
that many, if not most, of the absorption (and perhaps also emission)
line signatures seen in AGN and QSOs are formed in such accretion disk
winds. We therefore agree with Ganguly \& Brotherton (2008) that a 
comprehensive look at a wide range of outflow tracers is required in
order to develop a full empirical picture of these disk winds.
\nocite{hamann,murray,knigge,long}

The empirical distinctions between objects exhibiting different kinds
of outflow tracers are important clues in this process. For example, if
BALQSOs and other ALQSOs are literally ``the same thing viewed from
different angles'', it could be highly relevant that they occupy distinct
modes of the $\log{{\rm AI}}$\ distribution. For example, in the context of
orientation-based 
unification schemes, a restricted AI-range for BALQSOs would probably
imply that the BAL-forming region of the outflow has clearly
delineated physical boundaries. This would  
ensure that there is little room for overlap between sightlines looking
into this part of the outflow (and seeing a BAL) and sightlines
looking across it (and seeing only narrower absorption
features). However, this conclusion cannot yet be considered
robust, since different viable decompositions of the AI distribution 
can produce different AI-ranges for BALQSOs.

\section*{Acknowledgements}

We would like to thank Gordon Richards and Jonathan Trump for helpful
responses to several questions, as well as the anonymous referee for
an insightful and constructive report. 

This work is supported at the University of Southampton and the
University of Leicester by the Science and Technology Facilities
Council (STFC). 
    
Funding for the SDSS and SDSS-II has been provided by the Alfred P. Sloan
Foundation, the Participating Institutions, the National Science Foundation,
the U.S. Department of Energy, the National Aeronautics and Space
Administration, the Japanese Monbukagakusho, the Max Planck Society, and the
Higher Education Funding Council for England. The SDSS Web Site is
http://www.sdss.org/.

The SDSS is managed by the Astrophysical Research Consortium for the
Participating Institutions. The Participating Institutions are the American
Museum of Natural History, Astrophysical Institute Potsdam, University of
Basel, University of Cambridge, Case Western Reserve University, University of
Chicago, Drexel University, Fermilab, the Institute for Advanced Study, the
Japan Participation Group, Johns Hopkins University, the Joint Institute for
Nuclear Astrophysics, the Kavli Institute for Particle Astrophysics and
Cosmology, the Korean Scientist Group, the Chinese Academy of Sciences
(LAMOST), Los Alamos National Laboratory, the Max-Planck-Institute for
Astronomy (MPIA), the Max-Planck-Institute for Astrophysics (MPA), New Mexico
State University, Ohio State University, University of Pittsburgh, University
of Portsmouth, Princeton University, the United States Naval Observatory, and
the University of Washington.

\nocite{becker}
\nocite{dai}
\nocite{shankar}
\nocite{maddox}
\nocite{ganguly1}
\nocite{ganguly2}
\nocite{nestor}

\bibliographystyle{mn2e}

\begin{thebibliography}{}

\bibitem[\protect\citeauthoryear{Ashman, Bird, 
\& Zepf}{1994}]{ashman94} Ashman K.~M., Bird C.~M., Zepf S.~E., 1994, AJ, 108, 2348 

\bibitem[\protect\citeauthoryear{Becker et al.}{2001}]{becker} 
Becker, R.~H. et al. 2001, ApJS, 135, 227

\bibitem[\protect\citeauthoryear{Dai, Shankar, 
\& Sivakoff}{2008}]{dai} Dai X., Shankar F., Sivakoff G.~R., 2008, ApJ, 672, 108 

\bibitem[\protect\citeauthoryear{Di Matteo, Springel, 
\& Hernquist}{2005}]{DiMatteo} Di Matteo T., Springel V., Hernquist L., 2005, Natur, 433, 604 

\bibitem[\protect\citeauthoryear{Elvis}{2000}]{elvis} Elvis 
M., 2000, ApJ, 545, 63 

\bibitem[\protect\citeauthoryear{Foltz et al.}{1990}]{foltz90} 
Foltz C.~B., Chaffee F.~H., Hewett P.~C., Weymann R.~J., Morris S.~L., 
1990, BAAS, 22, 806 

\bibitem[\protect\citeauthoryear{Ganguly et al.}{2007}]{ganguly2} 
Ganguly, R. et al. 2007, ApJ, 665, 990

\bibitem[\protect\citeauthoryear{Ganguly \& Brotherton}{2008}]{ganguly1} 
Ganguly, R. \& Brotherton, M.~S. 2008, ApJ, 672, 102

\bibitem[\protect\citeauthoryear{Hall et al.}{2002}]{hall02} 
Hall P.~B., et al., 2002, ApJS, 141, 267 

\bibitem[\protect\citeauthoryear{Hamann, Korista, 
\& Morris}{1993}]{hamann} Hamann F., Korista K.~T., Morris S.~L., 1993, ApJ, 415, 541 

\bibitem[\protect\citeauthoryear{Hewett 
\& Foltz}{2003}]{hewett03} Hewett P.~C., Foltz C.~B., 2003, AJ, 125, 1784 

\bibitem[\protect\citeauthoryear{King}{2003}]{king} King A., 
2003, ApJ, 596, L27 

\bibitem[\protect\citeauthoryear{Knigge, Woods, 
\& Drew}{1995}]{knigge} Knigge C., Woods J.~A., Drew J.~E., 1995, MNRAS, 273, 225 

\bibitem[\protect\citeauthoryear{{Kohonen}}{{Kohonen}}{2001}]{kohonen01}
{Kohonen} T.,  2001, {Self-organizing maps}.
Self-organizing maps.3rd ed.Berlin: Springer, 2001, xx, 501 p.Springer series
  in information sciences, ISBN 3540679219

\bibitem[\protect\citeauthoryear{Long 
\& Knigge}{2002}]{long} Long K.~S., Knigge C., 2002, ApJ, 579, 725 

\bibitem[\protect\citeauthoryear{Maddox etl.}{2008}]{maddox}
Maddox, N, Hewett, P.~C., Warren, S.~J., Croom, S.~M. 2008, MNRAS, in press (arXiv:0802.3650)

\bibitem[\protect\citeauthoryear{Murray et al.}{1995}]{murray} 
Murray N., Chiang J., Grossman S.~A., Voit G.~M., 1995, ApJ, 451, 498 

\bibitem[\protect\citeauthoryear{Nestor, D., Hamann, F. \& Rodriguez
Hidalgo, P.}{2008}]{nestor} Nestor, D., Hamann, F. \& Rodriguez
  Hidalgo, P. 2008, MNRAS, in press (arXiv:0803:0326)

\bibitem[\protect\citeauthoryear{North, Knigge, 
\& Goad}{2006}]{north06} North M., Knigge C., Goad M., 2006, MNRAS, 365, 1057 

\bibitem[\protect\citeauthoryear{Reichard et 
al.}{2003}]{reichard03b} Reichard T.~A., et al., 2003, AJ, 126, 
2594 

\bibitem[\protect\citeauthoryear{Richards et 
al.}{2002}]{Richards} Richards G.~T., et al., 2002, AJ, 123, 
2945 

\bibitem[\protect\citeauthoryear{Scannapieco, Silk, 
\& Bouwens}{2005}]{scannapieco} Scannapieco E., Silk J., Bouwens R., 2005, ApJ, 635, L13 

\bibitem[\protect\citeauthoryear{Schneider et 
al.}{2005}]{schneider} Schneider D.~P., et al., 2005, AJ, 130, 
367 

\bibitem[\protect\citeauthoryear{Shankar, Dai, \& Sivakoff}{2008}]{shankar} 
Shankar F., Dai, X., Sivakoff G.~R., 2008, ApJ, submitted (arXiv:0801.4379)

\bibitem[\protect\citeauthoryear{Silk 
\& Rees}{1998}]{silk} Silk J., Rees M.~J., 1998, A\&A, 331, L1 

\bibitem[\protect\citeauthoryear{Stocke et al.}{1992}]{stocke92} 
Stocke J.~T., Morris S.~L., Weymann R.~J., Foltz C.~B., 1992, ApJ, 396, 487 

\bibitem[\protect\citeauthoryear{Tolea, Krolik, 
\& Tsvetanov}{2002}]{Tolea} Tolea A., Krolik J.~H., Tsvetanov Z., 2002, ApJ, 578, L31 

\bibitem[\protect\citeauthoryear{Trump et al.}{2006}]{trump06} 
Trump J.~R., et al., 2006, ApJS, 165, 1 

\bibitem[\protect\citeauthoryear{Weymann et 
al.}{1991}]{weymann91} Weymann R.~J., Morris S.~L., Foltz C.~B., 
Hewett P.~C., 1991, ApJ, 373, 23 

\bibitem[\protect\citeauthoryear{{Wyszomirski}}{{Wyszomirski}}{1992}]{wys}
{Wyszomirski} T.,  1992, J. Theor. Biol., 158, 109

\end{thebibliography}

\label{lastpage}

%% for reference, the list in ads format
%% 1994AJ....108.2348A
%% 2008ApJ...672..108D
%% 2005Natur.433..604D
%% 1990BAAS...22..806F
%% 2002ApJS..141..267H
%% 1993ApJ...415..541H
%% 2003AJ....125.1784H
%% 2003ApJ...596L..27K
%% 1995MNRAS.273..225K
%% 2002ApJ...579..725L
%% 1995ApJ...451..498M
%% 2006MNRAS.365.1057N
%% 2003AJ....126.2594R
%% 2002AJ....123.2945R
%% 2005ApJ...635L..13S
%% 2005AJ....130..367S
%% 1998A&A...331L...1S
%% 1992ApJ...396..487S
%% 2002ApJ...578L..31T
%% 2006ApJS..165....1T

\end{document}